\documentclass[pra,twocolumn,showpacs,preprintnumbers,amsmath,amssymb,superscriptaddress]{revtex4}

\usepackage{graphicx}
\usepackage{natbib}
\usepackage{latexsym}
\begin{document}

\newcommand{\ei}{\hat{a}}
\newcommand{\eidag}{\hat{a}^{\dag}}
\newcommand{\hn}{\hat{n}}

\newcommand{\ddt}{\frac{d}{dt}}
\newcommand{\Lx}{\hat{L}_x}
\newcommand{\Ly}{\hat{L}_y}
\newcommand{\Lz}{\hat{L}_z}
\newcommand{\Li}{\hat{L}_i}
\newcommand{\Lj}{\hat{L}_j}
\newcommand{\Lk}{\hat{L}_k}
\newcommand{\Lp}{\hat{L}_+}
\newcommand{\Lm}{\hat{L}_-}
\newcommand{\Lpm}{\hat{L}_\pm}

\newcommand{\Dxx}{\Delta_{xx}}
\newcommand{\Dyy}{\Delta_{yy}}
\newcommand{\Dzz}{\Delta_{zz}}
\newcommand{\Dxy}{\Delta_{xy}}
\newcommand{\Dyz}{\Delta_{yz}}
\newcommand{\Dzx}{\Delta_{zx}}

\title{Decoherence and entanglement in a bosonic Josephson junction: Bose-enhanced quantum-Zeno control of phase-diffusion}

\author{Y. Khodorkovsky\footnote{Present address: Department of Chemical Physics, The Weizmann Institute of Science, Rehovot 76100, Israel.}}
\affiliation{Department of Chemistry, Ben-Gurion University of
the Negev, P.O.B. 653, Beer-Sheva 84105, Israel}
\author{G. Kurizki}
\affiliation{Department of Chemical Physics, The Weizmann Institute of Science, Rehovot 76100, Israel}
\author{A. Vardi}
\affiliation{Department of Chemistry, Ben-Gurion University of
the Negev, P.O.B. 653, Beer-Sheva 84105, Israel}

\begin{abstract}
We study the effect of decoherence on dynamical phase diffusion in the two-site Bose-Hubbard model. Starting with an odd parity excited coherent state, the initial loss of single particle coherence varies from small bound oscillations in the Rabi regime, through hyperbolic depletion in the Josephson regime, to a Gaussian decay in the Fock regime. The inclusion of local-site  noise, measuring the relative number difference between the modes, is shown to enhance phase-diffusion. In comparison, site-indiscriminate noise measuring the population imbalance between the two quasi-momentum modes, slows down  the loss of single-particle coherence. Decoherence thus either enhances or suppresses phase-diffusion, depending on the details of system-bath coupling and the overlap of decoherence pointer states with collisional-entanglement pointer states. The deceleration of phase-diffusion due to the coupling with the environment may be viewed as a many-body quantum-Zeno effect. The extended effective decay times in the presence of projective measurement, are further enhanced with increasing number of particles $N$, by a bosonic factor of $\sqrt{N}$ in the Fock regime and $N/\log{N}$ in the Josephson regime.  
\end{abstract}  
\maketitle

\section{\label{Sec:intro}Introduction}

Decoherence and entanglement constitute the last frontiers of quantum mechanics. The crucial role of decoherence in the quantum-classical transition, was gradually revealed over the past three decades \cite{Zurek03,Schlosshauer04,Joos,Schlosshauer}. According to this 'environmentally induced superselection' scenario \cite{Zeh70,Zeh73,Zurek81,Zurek82}, the environment interactions with an open quantum system, select a preferred basis set of so called pointer states that are immune to decoherence. Pointer states define realizable observables, thereby restricting the number of meaningful measurements to a handful of classical quantities. For a multipartite system, all correlations but those existing in the pointer states, are rapidly destroyed by decoherence, so that out of all possible decompositions of the mixed multipartite state into a statistical mixture of pure states, only the decomposition into pointer states is resilient. Hence, after a short time from its onset, decoherence drives the system into an incoherent sum of eigenstates of {\it particular} classical observables.

The notion of classicality is fundamental to the theory of quantum Bose gases \cite{Pitaevskii}. In fact, the very definition of Bose-Einstein condensation (BEC) is the emergence of single-particle coherence \cite{Penrose56}, which enables the Gross-Pitaevskii (GP) classical field description of the interacting many-particle BEC state, 
\begin{equation}
\label{GP}
i \hbar \frac{\partial}{\partial t} \psi(\mathbf{r},t) = 
\left(- \frac{\hbar^2 \mathbf{\nabla}^2}{2m} + 
V_{ext}(\mathbf{r}) + 
U \left|\psi(\mathbf{r},t)\right|^2 \right) \psi(\mathbf{r},t)~. 
\end{equation}
In the above, $\psi(\mathbf{r},t)$ is the classical condensate order parameter, $m$ is particle's mass, $V_{ext}(\mathbf{r})$ is an external potential (e.\ g.\ an harmonic trap), and $U=4 \pi \hbar^2 a/m$ is the two body interaction strength (set by low-energy $s$-wave scattering with characteristic scattering length $a$). Thus, somewhat ironically, the epitomization of quantum mechanics on a mesoscopic scale, is in fact a classical, uncorrelated state. 

At zero temperature, the magnitude of quantum fluctuations around the condensate state, scales down as $1/\sqrt{N}$, with increasing particle number $N$. Since $N$ in dilute BEC experiments is in the range of $10^4-10^7$ particles, quantum fluctuations are characteristically small and the classical field description is an excellent approximation. However, the initially small value of quantum fluctuations does not necessarily guarantee that they will remain small away from equilibrium. In particular, if the classical dynamics is unstable, fluctuations grow rapidly and become significant on timescales that only grow logarithmically with $N$ \cite{Zurek03,CastinDum97,Vardi01,Anglin01,Khodor08,Boukobza09}. Single-particle coherence is thus dynamically lost due to interparticle entanglement caused by the interactions, at a rate which depends on the stability of the classical dynamics. For large $N$, the early stages of this process may be viewed as the decoherence of the reduced single particle state due to its coupling to a bath of quantum correlations \cite{Vardi01}. As such, collisional dephasing will have its own set of pointer states, untouched by the coupling to the  'environment' of quantum correlations.

Coupling the many-body system to an external bath (e.g. at finite temperature, where the bath consists of the thermal-cloud degrees of freedom), we are confronted with an interesting situation, where the decoherence of the system due to its entanglement with the bath, may affect the loss of single-particle coherence due to the internal entanglement of different particles, thereby affecting its classicality. In previous work we found that local site noise enhances the collisional loss of single-particle coherence in a two mode BEC, also known as  {\it phase-diffusion} \cite{Vardi01,Anglin01,Khodor08,Boukobza09,Castin97,Lewenstein96,Wright96,Javanainen97,
GreinerPD02,Jo07,Widera08}. Hence such noise drives the system away from classical behavior .   

Here we provide the details of recent work \cite{Khodor08} showing that decoherence can also {\it suppress} the loss of single-particle coherence in the interacting two-mode system. Since inelastic collisions measure the relative number between the modes, the pointer states of collisional dephasing are relative-number (Fock) states. Starting from an antisymmetric coherent state in the absence of an external bath, it dephases in a rate which depends on the coupling parameter $\kappa=UN/J$, where $J$ is the hopping energy between the modes \cite{Boukobza09}. Subjecting the same state to different forms of decoherence, we find that when the pointer states of the external bath coincide with the relative number states, dephasing is enhanced. On the other hand, when the pointer states have a definite number between the odd- and even superpositions of the two modes, so as to include the initial state, dephasing is slowed down. Since the amplification of quantum noise is sensitive to the relative phase between the odd- and even superpositions of the modes, the suppression of single-particle dephasing can be viewed as a quantum Zeno effect (QZE) \cite{Khalfin68,Misra77,Itano90,Kofman00,Kofman01,Streed06,Haroche06}, resulting from the continuous projection onto relative number states which prevents this phase from taking a definite value. The predicted suppression of phase-diffusion is shown to depend on the number of particles as $N/\log{N}$ in the $1<\kappa<N^2$ Josephson regime and as $\sqrt{N}$ in the $\kappa>N^2$ Fock regime, making the QZE dramatically more pronounced for high $N$. We note that another example of increased coherence due to the combined effects of lossy dissipation and interactions, has been found in the bimodal BEC system in the form of a stochastic resonance \cite{Witthaut08,Witthaut09}.

The model system is presented in Sec.~II and collisional dephasing is discussed in Sec.~III, demonstrating different qualitative behavior in three distinct coupling regimes. In Sec.~IV, we open the system and study the influence of two decoherence mechanisms on the phase-diffusion  process. Summary and conclusions are presented in Sec.~V.

\section{The Two-Site Bose-Hubbard Model and Dynamical Equations}

Starting out as a bare-bones toy model \cite{Javanainen86,Javanainen96,Jack96,Milburn97,Parkins98,Ruostekovski98,Smerzi97,Zappata98,Villain99,Shchesnovich08}, which retains many of the qualitative features of the many-mode Bose-Hubbard model \cite{Fisher89,Jaksch98,Greiner02,Spielman07}, the two-mode BEC was recently shown to be useful in the description of quantum interference experiments \cite{Schumm05,Hofferberth07,Shin05,JoChoi07}, bosonic Josephson junctions \cite{Albiez05,ShinJo05,Gati06,Gati07,Levy07} and quantum tunneling in an array of separated double wells \cite{Anderlini06,Sebby-Strabley07,Folling07}. The tight-binding condition for such wells is
\begin{equation}
|U|n=\frac{4\pi\hbar^2|a|}{m}\frac{N}{\frac{4}{3}\pi l^3} \ll \hbar\omega=\hbar\frac{\hbar}{ml^2}~,
\end {equation}
where $n$ is the number density,  $l=\sqrt{\hbar/(m\omega)}$ is the characteristic trap size and $\omega$ is the trap frequency in either well. Thus, provided that $l \gg 3 N|a|$, the system is accurately described by the quantized Josephson Hamiltonian \cite{Makhlin01},
\begin{equation}
\label{Ham}
\hat{H}=-\frac{J}{2}\left(\eidag_1 \ei_2 + \ei_1 \eidag_2 \right)
-\frac{\Delta}{2}\left(\hn_1-\hn_2\right)+ \frac{U}{4}\left(\hn_1-\hn_2\right)^2, 
\end{equation}
where $\ei_i$ and $\eidag_i$ are the annihilation and creation operators respectively, for boson particles in the mode $i=1,2$ with corresponding particle number operators $\hn_{i}=\eidag_{i}\ei_{i}$, $\Delta$ is a bias potential between the modes, which in the following is set to zero, and $J$ is the matrix element coupling the modes. In the above we have set $\hbar=1$,  and eliminated $c$-number terms proportional to the conserved total number of particles $N=\hn_1+\hn_2$.

The quantized Josephson model is easily mapped onto a spin problem by defining the three $SU(2)$ generators \cite{Vardi01,Anglin01,Khodor08,Boukobza09}
\begin{eqnarray}
\label{Lx}
\Lx&=&\frac{\eidag_1 \ei_2 + \ei_1 \eidag_2}{2}~,\\
\label{Ly}
\Ly&=&\frac{\eidag_1 \ei_2 - \ei_1 \eidag_2}{2i}~,\\
\label{Lz}
\Lz&=&\frac{\hn_1 - \hn_2}{2}~,
\end{eqnarray}
which determine the reduced single-particle density matrix (SPDM)
\begin{eqnarray}
\label{SPDM}
R_{ij}&=&\frac{1}{N} \langle \eidag_i \ei_j \rangle \\
~&=&\frac{1}{N} \left( \frac{1}{2} \langle \hat{N} \rangle \mathbf{1} +
\langle \Lx \rangle \boldsymbol{\sigma}_{\mathbf{x}} +
\langle \Ly \rangle  \boldsymbol{\sigma}_{\mathbf{y}}+
\langle \Lz \rangle  \boldsymbol{\sigma}_{\mathbf{z}} \right)_{ij}~,\nonumber
\end{eqnarray}
where $\mathbf{1},\boldsymbol{\sigma}_{\mathbf{x}},\boldsymbol{\sigma}_{\mathbf{y}},\boldsymbol{\sigma}_{\mathbf{z}}$ are the identity and Pauli matrices respectively, and $i,j=1,2$. With these definitions, the Hamiltonian (\ref{Ham}) assumes the form
\begin{equation}
\label{Ham_short}
\hat{H}=-J \Lx - \Delta \Lz + U\Lz^2~.
\end{equation}

\begin{figure}
\centering
\includegraphics[width=0.5\textwidth] {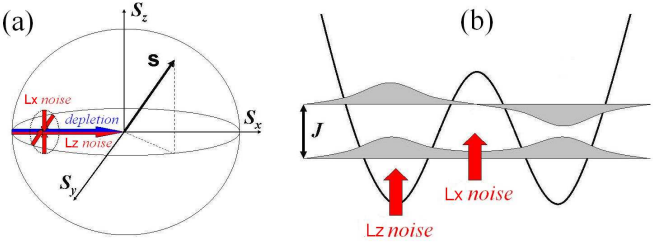}
\caption{(a) The $(-1,0,0)$ initial state on the Bloch sphere subjected to phase-diffusion, local $\Lz$ noise, and site-indiscriminate $\Lx$ noise. Phase-diffusion and local noise drive the Bloch vector  towards the $s_z$-axis, while the $\Lx$ noise projects it on the $s_x$-axis. (b) $\Lz$ noise distinguishes between the left and the right well states, while $\Lx$ noise distinguishes between the even and the odd quasi-momentum states.} 
\label{qzefig1}
\end{figure}  

As shown in Fig~\ref{qzefig1}, we account for the effect of two types of noise on the two-mode dynamics, by using the quantum kinetic Master equation,
\begin{equation}
\frac{d}{dt}\hat{\rho}=\frac{i}{\hbar}\left[\hat{\rho},\hat{H}\right]-\Gamma_z\left[\Lz,\left[\Lz,\hat{\rho}\right]\right]-\Gamma_x\left[\Lx,\left[\Lx,\hat{\rho}\right]\right]~,
\label{master}
\end{equation}
where $\hat{\rho}$ is the $N$-particle density operator. The $\Lz$ noise term on the r.h.s. of Eq. (\ref{master}) corresponds to the dephasing of the two local (quasi-coordinate) modes, as may be caused by collisions with thermal particles \cite{Anglin97}. The $\Lx$ noise term is the same for the phase between the odd- and even combinations of the site-mode functions (i.e. the two quasi-momentum modes). This site-indiscriminate noise may be implemented e.g. by a stochastic modulation of the potential barrier between the sites. Since the odd quasi-momentum mode has a node between the sites whereas the even quasi-momentum mode does not, such perturbation will serve to measure the population imbalance between them. Setting $\Delta=0$, the equations of motion for the angular momentum operators are
\begin{eqnarray}
\ddt \Lx&=&-U\left(\Ly \Lz + \Lz \Ly \right)-\Gamma_z\Lx~,\nonumber\\
\label{heisen}
\ddt \Ly&=&J \Lz + U \left(\Lz \Lx + \Lx \Lz \right)-(\Gamma_z+\Gamma_x)\Ly~, \nonumber\\
\ddt \Lz&=&-J \Ly-\Gamma_x\Lz~.
\end{eqnarray}

The classical limit of the two-mode model (\ref{Ham_short}) is obtained by taking the expectation values of Eqs. (\ref{heisen}) and using the Gaussian (coherent) approximation
\begin{equation}
\label{approx1}
\langle \Li \Lj \rangle \approx \langle \Li \rangle \langle \Lj \rangle~,
\end{equation}
which is equivalent to replacing the spin operators $\Li$ by classical $c$-numbers. Defining the single-particle Bloch vector
\begin{eqnarray}
\mathbf{s}=(s_x,s_y,s_z)&=&\left( \frac{2\langle \Lx \rangle}{N},\frac{2\langle \Ly \rangle}{N},\frac{2\langle \Lz \rangle}{N}\right) \\
~&=&\Bigl(2Re(R_{12}),2Im(R_{12}),R_{11}-R_{22}\Bigr)~, \nonumber 
\end{eqnarray}
rescaling time as $\tau=Jt$, and substituting into (\ref{heisen}), we obtain the Bloch equations:
\begin{eqnarray}
\dot{s}_x&=&-\kappa s_y s_z-\gamma_z s_x\nonumber\\
\label{bloch}
\dot{s}_y&=&s_z + \kappa s_z s_x-\left(\gamma_x+\gamma_z\right)s_y.\\
\dot{s}_z&=&-s_y-\gamma_x s_z,\nonumber
\end{eqnarray}
where $\dot s_i=ds_i/d\tau$, $\kappa=UN/J$ is the coupling parameter and $\gamma_i=\Gamma_i/J$ are the rescaled decoherence rates.

For $\gamma_x=\gamma_z=0$, the mean-field equations (\ref{bloch}) depict the unitary rotations of the Bloch vector with conserved norm $|\mathbf{s}|=1$. Beyond the mean-field approximation however, single-particle coherence will be lost due to entanglement between the particles, even at zero temperature where no noise is present. Since we are interested in the effect of phase-noise on this entanglement process, we must use a quantum formalism and probe its convergence to the classical (mean-field) limit with and without noise. A partial account of the dynamics of quantum fluctuations, is provided by the Bogoliubov Backreaction (BBR) formalism \cite{Vardi01,Anglin01,Tikhonenkov07}, based on the higher-order truncation scheme, 
\begin{eqnarray}
\langle \Li \Lj \Lk \rangle & \approx & \langle \Li \Lj \rangle \langle \Lk \rangle 
+ \langle \Li \rangle \langle \Lj \Lk \rangle + \langle \Li \Lk \rangle \langle \Lj \rangle \nonumber\\
\label{approx2}
~&~& -2 \langle \Li \rangle \langle \Lj \rangle \langle \Lk \rangle~.
\end{eqnarray}
Using approximation (\ref{approx2}) to truncate the hierarchy of dynamical equations for correlation functions, we obtain, 
\begin{eqnarray}
\dot{s}_x&=&-\kappa s_y s_z -\frac{\kappa}{2} \Dyz - \gamma_z s_x~,\nonumber\\
\dot{s}_y&=&s_z + \kappa s_z s_x + \frac{\kappa}{2}\Dzx-\left(\gamma_z+ \gamma_x\right)s_y~,\nonumber\\
\dot{s}_z&=&-s_y -\gamma_x s_z~,\nonumber\\
\dot{\Delta}_{xx}&=&-2\kappa\left(s_y \Dzx + s_z \Dxy \right)-2\gamma_z\left(\Dxx-\Dyy-2s_y^2\right)~,\nonumber\\
\dot{\Delta}_{yy}&=&2\left(1+\kappa s_x\right)\Dyz + 2\kappa s_z \Dxy \nonumber\\ 
~&~&-2\gamma_z\left(\Dyy-\Dxx-2s_x^2\right)-2\gamma_x\left(\Dyy-\Dzz-2s_z^2\right)~,\nonumber\\
\label{BBR}
\dot{\Delta}_{zz}&=&-2\Dyz - 2\gamma_x\left(\Dzz-\Dyy-2s_y^2\right)~,\\
\dot{\Delta}_{xy}&=&\left(1+\kappa s_x\right)\Dzx-\kappa s_y \Dyz + \kappa s_z \left(\Dxx-\Dyy\right)\nonumber \\
~&~&-4\gamma_z\left(\Dxy+s_x s_y\right)-\gamma_x\Dxy~,\nonumber\\
\dot{\Delta}_{yz}&=&\left(\Dzz-\Dyy\right)+\kappa\left(s_z\Dzx + s_x \Dzz\right)\nonumber\\
~&~&-\gamma_z\Dyz -4\gamma_x\left(\Dyz+s_y s_z \right)~,\nonumber\\
\dot{\Delta}_{zx}&=&-\Dxy -\kappa\left(s_z \Dyz + s_y \Dzz\right)-\left(\gamma_z+\gamma_x\right)\Dzx~,\nonumber
\end{eqnarray} 
where the two point normal correlation functions $\Delta_{ij}$ are defined as,
\begin{equation}
\Delta_{ij}=\frac{4}{N^2}\left( \langle \Li \Lj + \Lj \Li \rangle -2\langle \Li \rangle \langle \Lj \rangle \right)~.
\end{equation}

Finally, a full quantum solution may be obtained numerically in the Fock representation of mutual $\hat{{\bf L}}^2,\Lz$ eigenstates, 
\begin{equation}
\label{relative_Fock_basis}
\left|l,m\right\rangle=\frac{1}{\sqrt{\left(l+m\right)!\left(l-m\right)!}}\left(a_1^\dag\right)^{l+m}\left(a_2^\dag\right)^{l-m}|0\rangle, 
\end{equation}
satisfying,
\begin{eqnarray}
\hat{{\bf L}}^2\left|l,m\right\rangle&=&l(l+1)\left|l,m\right\rangle,\\
\Lz\left|l,m\right\rangle&=&m\left|l,m\right\rangle~,
\end{eqnarray}
with $m=-l,\dots,0,\dots,l$. The Hamiltonian (\ref{Ham_short}), as well as the operators producing the decoherence in the master equation (\ref{master}), commute with $\hat{{\bf L}}^2$ due to particle-number conservation. Consequently, the total angular momentum is fixed at $l=N/2$. We shall therefore denote the Fock states below as $\left|m\right\rangle\equiv\left|l=N/2,m\right\rangle$. Using this basis set of $N+1$ relative-number states to represent $\hat{\rho}$,
\begin{equation}
\hat{\rho}(t)=\sum_{m,m'=-l}^l \rho_{m,m'}(t)|m\rangle\langle m'|~,
\label{npdensity}
\end{equation}
 the master equation (\ref{master}) takes the form, 
\begin{widetext}
\begin{eqnarray}
\frac{d{\rho}_{mm'}}{d\tau}&=&-\frac{i}{2}\left[\sqrt{l(l+1)-m'(m'+1)}\rho_{m,m'+1}+\sqrt{l(l+1)-m'(m'-1)}\rho_{m,m'-1}\right.\nonumber\\
~&~&\left.-\sqrt{l(l+1)-m(m+1)}\rho_{m+1,m'}-\sqrt{l(l+1)-m(m-1)}\rho_{m-1,m'}\right]\nonumber\\
~&~&+\left[\frac{i\kappa}{N}\left({m'}^2-m^2\right)-\gamma_z(m-m')^2-\gamma_x\left(l(l+1)-\frac{m^2+{m'}^2}{2}\right)\right]\rho_{m,m'}\nonumber\\
~&~&-\frac{\gamma_x}{4} \left[\sqrt{\left(l(l+1)-(m+1)^2\right)^2-(m+1)^2}\rho_{m+2,m'}\right.\nonumber\\
~&~&+\sqrt{\left(l(l+1)-(m-1)^2\right)^2-(m-1)^2}\rho_{m-2,m'} \nonumber\\
~&~&+\sqrt{\left(l(l+1)-(m'+1)^2\right)^2-(m'+1)^2}\rho_{m,m'+2}\nonumber\\
~&~&\left.+\sqrt{\left(l(l+1)-(m'-1)^2\right)^2-(m'-1)^2}\rho_{m,m'-2}\right]\nonumber\\
~&~&+\frac{\gamma_x}{2}\left[\sqrt{\left(l(l+1)-m(m+1)\right)\left(l(l+1)-m'(m'+1)\right)}\rho_{m+1,m'+1}\right.\nonumber\\
~&~&+\sqrt{\left(l(l+1)-m(m-1)\right)\left(l(l+1)-m'(m'-1)\right)}\rho_{m-1,m'-1}\nonumber\\
~&~&+\sqrt{\left(l(l+1)-m(m+1)\right)\left(l(l+1)-m'(m'-1)\right)}
\nonumber\\
~&~&\left.\times\rho_{m+1,m'-1}+\sqrt{\left(l(l+1)-m(m-1)\right)\left(l(l+1)-m'(m'+1)\right)}\rho_{m-1,m'+1}\right]\label{masterelement}
\end{eqnarray}
\end{widetext}

The numerical solution of Eq. (\ref{masterelement}) will be used in the following sections to verify analytic predictions based on the linearization of the BBR equations (\ref{BBR}). 

\section{Phase Diffusion}
         
It is clear from the discussion in section II, that the classical states of the two-mode system coincide with the $SU(2)$ coherent states
\begin{eqnarray}  
|\theta,\phi\rangle&\equiv&
\exp\left(-i\phi\Lz\right)\exp\left(-i\theta\Ly\right) |-l\rangle\nonumber\\
\label{coherento}
~&=&\left[1+\tan^2\left(\frac{\theta}{2}\right)\right]^{-l}\nonumber\\
~&~&\times\sum_{m=-l}^l\left[\tan\left(\frac{\theta}{2}\right)e^{-i\phi}\right]^{l+m}
\left(
\begin{array}{c}
2l\\
l+m
\end{array}\right)^{1/2}|m\rangle\ ~.\nonumber
\end{eqnarray}
where $0<\theta<\pi$ and $0<\phi<2\pi$ are the rotation angles of the state $|m=-l\rangle$ corresponding to the point $\mathbf{s}=(0,0,-1)$ on the $SU(2)$ Bloch sphere. The mean-field factorization  (\ref{approx1}) for these states, is accurate to order $1/N$, becoming exact for large $N$. 

Interactions between particles lead to phase-diffusion \cite{Vardi01,Anglin01,Khodor08,Boukobza09,Castin97,Lewenstein96,Wright96,Javanainen97,GreinerPD02,Jo07,Widera08}. For repulsive interactions ($U>0$), the collisional loss of single-particle coherence due to interparticle entanglement exhibits particularly rich dynamics \cite{Vardi01,Anglin01,Khodor08,Boukobza09} around the excited mean-field eigenstate 
\begin{equation}
\left|\frac{\pi}{2},\pi\right\rangle=\frac{1}{2^l}\sum_{m=-l}^l (-1)^{l+m}
\left(
\begin{array}{c}
2l\\
l+m
\end{array}\right)^{1/2}|m\rangle~,
\label{oddstate}
\end{equation}
corresponding to all atoms in the excited quasi-momentum mode (and hence to a well-defined relative phase $\phi=\pi$). The Bloch vector affiliated with the state (\ref{oddstate}) is $\mathbf{s_0}=(-1,0,0)$. Subjecting it to interactions, one obtains different qualitative behavior of phase-diffusion in the three distinct interaction regimes  \cite{Boukobza09}:  (a) the weak-interaction Rabi regime $0<\kappa<1$ , (b) the strong-interaction Josephson regime $1<\kappa<N^2$, and (c) the extreme strong-interaction Fock regime $\kappa>N^2$.

In what follows, we quantitatively characterize the early stages of phase-diffusion in the three interaction regimes. Separating $\mathbf{s}(t)=\mathbf{s_0}+\mathbf{\delta s}(t)$ and linearizing the mean-field equations (\ref{bloch}) about this stationary classical point with $\gamma_x=\gamma_z=0$, we obtain
\begin{eqnarray}
\frac{d}{d\tau}\left[\begin{array}{c}
\delta s_x\\
\delta s_y\\
\delta s_z
\end{array}\right]&=&
\left[\begin{array}{ccc}
0&-\kappa s_{z0}&-\kappa s_{y0}\\
\kappa s_{z0}&0&1+\kappa s_{x0}\\
0&-1&0
\end{array}\right]
\left[\begin{array}{c}
\delta s_x\\
\delta s_y\\
\delta s_z
\end{array}\right]\nonumber\\
~&=&\left[\begin{array}{ccc}
0&0&0\\
0&0&1-\kappa\\
0&-1&0
\end{array}\right]
\left[\begin{array}{c}
\delta s_x\\
\delta s_y\\
\delta s_z
\end{array}\right]\label{linear}
\end{eqnarray}
resulting in the secular equation $\lambda(\lambda^2+1-\kappa)=0$ with the generally complex eigenvalues 
\begin{equation}
\lambda_0=0~,~\lambda_{\pm}=\mp i\sqrt{1-\kappa}
\label{frequencies}
\end{equation}
 and the corresponding nonorthogonal natural modes 
\begin{eqnarray}
\delta_0&=&\delta s_x\nonumber\\
\delta_\pm&=&\cos\Theta\delta s_y\pm i \sin\Theta\delta s_z,
\end{eqnarray}
where $\tan\Theta=\sqrt{1-\kappa}$ (under this definition, the angle $\Theta$ and its trigonometric functions are real in the Rabi regime and imaginary in the Josephson and Fock regimes, where they may be replaced by the real angle $i\Theta$ and its real hyperbolic functions). For  $U>0$ and $\kappa<1$ (i.e. $0<\Theta<\pi/4$), all the characteristic frequencies are real, indicating the stability of the point $\mathbf{s_0}$. Classical fluctuation remain bound as, 
\begin{eqnarray}
\delta_+(\tau)&=&\delta_+(0)\exp({-i\lambda\tau}),\nonumber\\
\delta_-(\tau)&=&\delta_-(0)\exp({+i\lambda\tau}),
\label{stab_fluct_one}
\end{eqnarray}
with $\lambda=|\sqrt{1-\kappa}|$. Transforming back to $\delta s_i$, we obtain that fluctuations evolve like the sine and cosine of $\lambda\tau$,
\begin{eqnarray}
\delta s_x(\tau)&=&\delta s_x(0)\\
\delta s_y(\tau)&=&\delta s_y(0)\cos(\lambda\tau)+\delta s_z(0)\tan\Theta\sin(\lambda\tau)~,\nonumber\\
\delta s_z(\tau)&=&-\delta s_y(0)\cot\Theta\sin(\lambda\tau)+\delta s_z(0)\cos(\lambda\tau)~~,\nonumber
\label{stab_fluct_two}
\end{eqnarray}
so that 
\begin{equation}
\delta_+\delta_-=|\delta_+|^2=|\delta_-|^2=\cos^2\Theta\delta s_y^2+\sin^2\Theta\delta s_z^2,
\label{stable_conserv}
\end{equation}
is a constant of motion. In contrast, for $U>0$ and $\kappa>1$, imaginary characteristic frequencies appear and the point $\mathbf{s_0}$ becomes dynamically unstable,
\begin{eqnarray}
\delta_+(\tau)&=&\delta_+(0)\exp({+\lambda\tau}),\nonumber\\
\delta_-(\tau)&=&\delta_-(0)\exp({-\lambda\tau}).
\label{fluct_one}
\end{eqnarray} 
Consequently we obtain,
\begin{eqnarray}
\delta s_x(\tau)&=&\delta s_x(0)\label{fluct_two}\\
\delta s_y(\tau)&=&\delta s_y(0)\cosh(\lambda\tau)+\delta s_z(0)\tanh(i\Theta)\sinh(\lambda\tau)~,\nonumber\\
\delta s_z(\tau)&=&\delta s_z(0)\cosh(\lambda\tau)+\delta s_y(0)\coth(i\Theta)\sinh(\lambda\tau)~~,\nonumber
\end{eqnarray}
with $\tanh(i\Theta)=i\tan\Theta=\sqrt{\kappa-1}$ and $\coth(i\Theta)=-i\cot\Theta=1/\sqrt{\kappa-1}$. Thus, while in the Rabi regime the magnitude of $\delta_+ $ and $\delta_-$ is conserved separately, only the product $\delta_+\delta_-$ is conserved in the Josephson and Fock  regimes.  

The classical stability analysis of mean-field trajectories is invariably associated with the dynamics of quantum corrections. In order to track down the initial growth of these fluctuations, we linearize the BBR equations (\ref{BBR}) with $\gamma_x=\gamma_z=0$, about the point $(-1,0,0)$, to obtain the block-diagonal form
\begin{widetext}
\begin{eqnarray}
\frac{d}{d\tau}{\left[\begin{array}{c}
\Dxx\\
\Dyy\\
\Dzz\\
\Dyz\\
\Dxy\\
\Dzx
\end{array}\right]}&=&
\left[\begin{array}{ccc}
0&~&~\\
~&\begin{array}{ccc}
0&0&2(1-\kappa)\\
0&0&-2\\
-1&1-\kappa&0\end{array}&~\\
~&~&\begin{array}{cc}
0&1-\kappa\\
-1&0
\end{array}
\end{array}\right]
\left[\begin{array}{c}
\Dxx\\
\Dyy\\
\Dzz\\
\Dyz\\
\Dxy\\
\Dzx
\end{array}\right]~.
\label{BBRlin}
\end{eqnarray}
\end{widetext}
In addition to the conserved variance $\Dxx$, the linearized equations (\ref{BBRlin}) have the pertinent natural modes,
\begin{eqnarray}
\Delta_0&=&\cos^2\Theta\Dyy+\sin^2\Theta\Dzz\\
\Delta_\pm&=&\cos^2\Theta\Dyy-\sin^2\Theta\Dzz\pm i\sin(2\Theta)\Dyz,\nonumber 
\end{eqnarray} 
with the respective characteristic frequencies $0,2\lambda_\pm$. Substituting the initial values $\Dyy=\Dzz=2/N$, $\Dxy=\Dzx=\Dyz=\Dxx=0$ of quantum fluctuations for the coherent state $|\pi/2,\pi\rangle$, we have
\begin{equation}
\Delta_0(\tau)=\frac{2}{N}~,~\Delta_\pm(\tau)=\frac{2}{N}\cos(2\Theta) e^{2\lambda_\pm\tau}. 
\label{BBReigendyn}
\end{equation}
Therefore the variances $\Delta_\pm$ are conserved separately in the Rabi regime, whereas the $\Delta_{ii}$ variances carry out small oscillations,
\begin{eqnarray}
\label{diagfluct}
\Dxx(\tau)&=&0,\\
\Dyy(\tau)&=&\frac{2\Delta_0+\left(\Delta_+ +\Delta_-\right)}{4\cos^2\Theta}=\frac{1+\cos(2\Theta)\cos(2\lambda\tau)}{N\cos^2{\Theta}},\nonumber\\
\Dzz(\tau)&=&\frac{2\Delta_0-\left(\Delta_+ +\Delta_-\right)}{4\sin^2\Theta}=\frac{1-\cos(2\Theta)\cos(2\lambda\tau)}{N\sin^2{\Theta}}.\nonumber
\end{eqnarray}
Since $\cos(2\Theta)>0$ for $0<\Theta<\pi/4$, relative-phase fluctuations $\Dyy$ are reduced at the expense of increased relative-number fluctuations $\Dzz$. The relative-phase squeezing of the excited state $|\pi/2,\pi\rangle$ is complementary to the well-known relative-number squeezing of the ground state $|\pi/2,0\rangle$ \cite{Orzel01}. 

\begin{figure}
\centering
\includegraphics[width=0.45\textwidth]{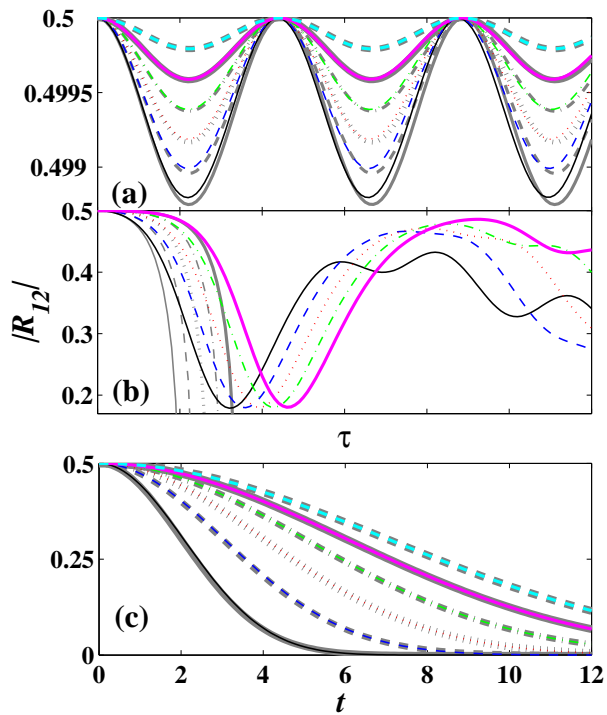}
\caption{(color online) Absolute value of single particle coherence as a function of rescaled time, starting from the coherent state $|\pi/2,\pi\rangle$ with (a) $\kappa=0.5$, (b) $\kappa=2$, and (c) $\kappa=\infty$ (i.e. $UN=2,J=0$). Solid black, dashed blue, dotted red, dash-dotted green, bold solid magenta and bold dashed cyan correspond to growing $N$ in all three figures, with $N=100, 120, 150, 200, 300, 600$ in (a), $N=50, 100, 200, 400, 800$ in (b), and $N=16, 36, 64, 100, 144, 196$ in (c). Gray lines correspond to the analytic forms of Eq. (\ref{Rstable}) in (a), Eq.\ (\ref{Runstable}) in (b), and Eq. (\ref{gauss}) in (c).} 
\label{qzefig2}
\end{figure} 

In order to relate the dynamics of correlation functions in (\ref{diagfluct}) to the loss of fringe-visibility in matter-wave interference experiments, we note that in the absence of decoherence ($\gamma_x=\gamma_z=0$), the BBR equations (\ref{BBR}) conserve the sum ${\bf s}^2+\left(\Dxx+\Dyy+\Dzz\right)/2$, which for the initial coherent state is $1+2/N$. The off-diagonal coherence of the reduced SPDM is thus,
\begin{eqnarray}
|R_{12}(\tau)|^2&=&\frac{1}{N^2}|\langle\Lp\rangle|^2=\frac{1}{N^2}\left(\langle \Lx\rangle^2+\langle \Ly\rangle^2\right)\nonumber\\
~&=&\frac{{\bf s}^2-s_z^2}{4}=\frac{1}{4}\left(1+\frac{2}{N}-\frac{\Dxx+\Dyy+\Dzz}{2}\right)\nonumber\\
~&=&\frac{1}{4}-\frac{\cot^2(2\Theta)}{N}\sin^2(\lambda\tau),
\label{Rstable}
\end{eqnarray}
where $\Lp=\Lx+i\Ly=\eidag_1\ei_2$. The fringe-visibility in the weak-interaction regime remains high and carries small oscillations whose amplitude is inversely proportional to the number of particles $N$. The two modes are {\it phase-locked} in the Rabi regime.

Repeating this process for strong interactions $\kappa>1$, the classical instability is associated with a rapid divergence from mean-field theory. This deviation is manifested in the squeezing of the initially coherent state according to Eq. (\ref{BBReigendyn}), by exponential reduction of the $\Delta_-$ variance at the expense of increasing  $\Delta_+$. Both $\Dyy$ and $\Dzz$ are amplified hyperbolically and the single-particle coherence is lost as,
\begin{equation}
|R_{12}(\tau)|^2=\frac{1}{4}-\frac{\coth^2(2i\Theta)}{N}\sinh^2(\lambda\tau)~.
\label{Runstable}
\end{equation} 
The phase-diffusion rate is therefore independent of the number of particles, but its onset time scales logarithmically with $N$ \cite{Vardi01,Anglin01}. The growth of purely ${\it quantum}$ correlations, related to the collisional entanglement between particles in the absence of an external bath, will thus vary depending on the stability of the ${\it classical}$ dynamics, determined by the value of the coupling parameter $\kappa$.

In Fig.~\ref{qzefig2} we plot the numerically calculated magnitude of the single-particle coherence $|R_{12}|$, as a function of the rescaled time $\tau$, for three values of $\kappa$, corresponding to the different interaction regimes. All the results in this figure are obtained for $\gamma_z=\gamma_x=0$. In the Rabi regime ($\kappa<1$), represented in Fig.~\ref{qzefig2}(a) by the value $\kappa=0.5$, all characteristic frequencies are real, the relative phase is locked, and the fringe-visibility exhibits the anticipated stable small oscillations, in excellent agreement with Eq. (\ref{Rstable}). The oscillation amplitude decreases reciprocally with increasing $N$ as $N\rightarrow\infty$, keeping $\kappa=UN/J$ fixed. Therefore in the Rabi regime, small ${\cal O}(1/N)$ initial quantum fluctuations remain small compared to the ${\cal O}(1)$ classical mean-field. 

\begin{figure}
\centering
\includegraphics[width=0.5\textwidth]{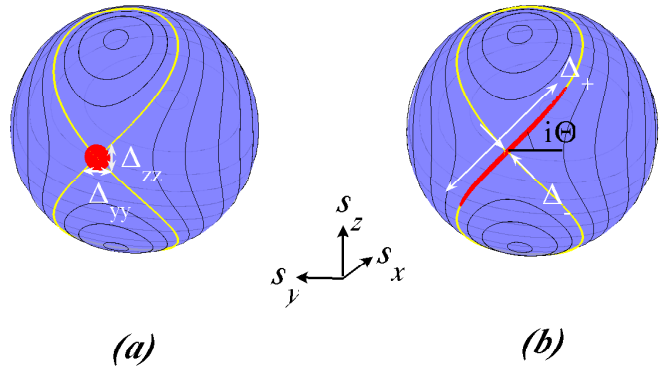}
\caption{(color online) Propagation of mean-field equations (\ref{bloch}) for $\gamma_x=\gamma_z=0$, with a distribution of initial conditions around the unstable point $\mathbf{s}=(-1,0,0)$. The initial distribution is shown in (a). The distribution at $\tau=2$ is shown in (b) } 
\label{qzefig3}
\end{figure}

\begin{figure}
\centering
\includegraphics[width=0.5\textwidth] {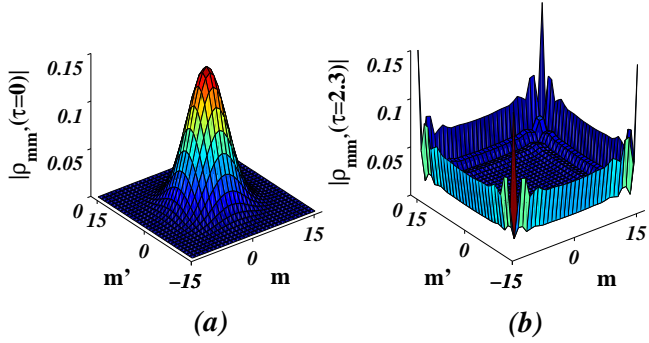}
\caption[Distribution of density matrices for dynamics without quantum noise]{Absolute values of the full density matrix entries in relative Fock basis, for exact evolution with $N=30$ and $\kappa=2$, without noise. (a) Initial coherent state $|\frac{\pi}{2},\pi\rangle$ at $\tau=0$. (b) The creation of the cat state at $\tau=2.3$, approximately given by $(|m=l\rangle + |m=-l\rangle)/\sqrt{2}$ (disregarding the relative phase).} 
\label{qzefig4}
\end{figure} 

By comparison, correlations grow rapidly in the Josephson regime. In Fig.~\ref{qzefig2}(b) we plot the evolution of single particle coherence, starting from the state (\ref{oddstate}), for $\kappa=2$. Since the characteristic frequencies of the linearized equations become imaginary in this regime, $\Delta_+$ fluctuations are amplified at the expense of attenuating $\Delta_-$, as described in Eq. (\ref{BBReigendyn}), and the initially coherent state is squeezed. The relation between this squeezing and the classical dynamical instability is illustrated in Fig.~\ref{qzefig3}, where the initial quantum propagation is approximated by mean-field evolution with a distribution of classical initial conditions which replace quantum fluctuations. While not as accurate as a complete stochastic calculation, this crude form of the truncated Wigner approach \cite{Santagiustina98, Gatti97} clearly shows how the classical saddle-point leads to the dynamical squeezing of quantum fluctuations and the amplification of quantum corrections as expected from Eq. (\ref{Runstable}). Since the deviation from mean-field classicality is proportional to $\sinh^2(\lambda\tau)/N$, the characteristic quantum-break time $\tau_{br}$ (the time at which this deviation crosses some predetermined threshold of ${\cal O}(1)$),  only grows logarithmically  with $N$, leading to significant phase-diffusion at short times, even for a large particle number. We note that this behavior differs from the phase-locking of the ground ($\phi=0$) coherent state in the Josephson regime \cite{Boukobza09}. The hyperbolic dependence of Eq. (\ref{Runstable}) gives an accurate description of the depletion at short times and a good estimate for the quantum break-time. At longer times, depletion becomes sufficiently large so that the linearized BBR equations (\ref{BBRlin}) are no longer adequate to describe the dynamics. The full nonlinear BBR equations however, remain valid and trace out the loss of single-particle coherence with good accuracy, as shown in Fig.~\ref{qzefig6}. Partial and full revivals are observed due to the finite number of ($N+1$) phase-space dimensions.

It is also instructive to explore phase-diffusion in the Josephson regime beyond the semiclassical squeezing picture. In Fig~\ref{qzefig4} we plot the entries of the $N$-particle density matrix ${\hat \rho}$ with $N=30$, for the initial coherent state (\ref{oddstate}) and for the evolved state at $\tau=2.3$. Interestingly, in the Josephson regime phase-diffusion of the odd $|\pi/2,\pi\rangle$ coherent state with repulsive interactions, results in the generation of a macroscopic Schr\"odinger's cat state (a macroscopic superposition of the two Fock states with all particles occupying either mode).  The same observation was made for the even coherent state $|\pi/2,0\rangle$, evolving into a macroscopic superposition state under phase diffusion with attractive interactions \cite{Louis01,Micheli03,Huang06}.

In the Fock regime, tunneling is effectively turned off and the number distribution is locked at its initial Gaussian form. Time evolution is thus restricted to phase-oscillations of the various Fock amplitudes. In this case, approximate analytical expressions may be found for the loss of multi-shot fringe visibility. Setting $J=0$, the time evolution of the initial coherent state due to the purely collisional Hamiltonian is
\begin{eqnarray}
\left|\psi(t)\right\rangle&=&\exp(-i\hat{H}t)\left|\frac{\pi}{2},\pi\right\rangle=\exp(-iU\Lz^2t)\left|\frac{\pi}{2},\pi\right\rangle\nonumber\\
~&=&\sum_{m=-l}^{l}\exp\left(-iU m^2 t\right)\Bigl|m\Bigr\rangle\left\langle m\Big{|}\frac{\pi}{2},\pi\right\rangle,
\end{eqnarray}
with the binomial coefficients of Eq. (\ref{oddstate}),
\begin{equation}
c_{m0}\equiv\left\langle m\Big{|}\frac{\pi}{2},\pi\right\rangle=
\frac{(-1)^{l+m}}{2^l}\left(
\begin{array}{c}
2l\\
l+m
\end{array}\right)^{1/2}~.
\end{equation}
The expectation value of $\Lp$ is thus
\begin{eqnarray}
\left\langle\Lp\right\rangle&=&\sum_{m',m=-l}^{l}c_{m'0}^*c_{m0}e^{iU\left(m'^2-m^2\right)t}\left\langle m'\right|\Lp\left|m\right\rangle\label{gauss_sum}\\
~&=&\sum_{m=-l}^{l} c_{(m+1)0}^*c_{m0}e^{iU(2m+1)t}\sqrt{(l-m)(l+m+1)}\nonumber~.
\end{eqnarray}
Since relative-number fluctuations  for the state $|\pi/2,\pi\rangle$ are Poissonian, we should only consider non-vanishing $c_{m0}$ with $m$ smaller than a few $\sqrt{N}=\sqrt{2l}$ standard deviations, so that for large $l$ (i.e. large $N$) we have $m,m+1\ll l$ for all relevant $m$. Therefore, we can replace $\sqrt{(l-m)(l+m+1)}\approx l=N/2$. Moreover, for large $l$ we can replace $c_{(m+1)0}\approx c_{m0}$, so that,
\begin{equation}
\left\langle\Lp\right\rangle\approx l e^{iUt}\sum_{m=-l}^{l} |c_{m0}|^2 e^{2iUmt}~.
\end{equation}
Thus, for large $N$, the time evolution of the single-particle coherence is just the Fourier transform of the initial number distribution. With the standard large-number replacement of the binomial distribution by a normal (Gaussian) distribution of the relative number difference with width $\Delta N_r=\sqrt{N}$, we finally obtain,
\begin{equation}
\label{gauss}
|R_{12}(\tau)|=\frac{1}{2} \exp\left(-\frac{U^2 N}{2J^2} \tau^2\right)~,
\end{equation}
with the characteristic dimensionless decay time of $J/(U\sqrt{N})=\sqrt{N}/\kappa$. This behavior is illustrated in Fig.~\ref{qzefig2}(c), where the gray lines corresponding to Eq. (\ref{gauss}) coincide with the numerical calculations. The same dependence of the single-particle coherence decay time on $(d\mu/dN_i)_{N_i=N/2}\Delta N_r$ (in our case $(d\mu/dN_i)_{N_i=N/2}=U$ and $\Delta N_r=\sqrt{N}$) was recently used to demonstrate squeezing via interference experiments \cite{Jo07}. Sub-Poissonian number fluctuations with $\Delta N_r=\sqrt{N}/s$ ($s$ being the squeezing parameter) resulted in slower relative-coherence loss during a purely collisional hold period, than would be expected from a coherent state with $\Delta N_r=\sqrt{N}$. Gaussian  decoherence was also predicted for the case of a single spin coupled to an environment of $N$ other distinguishable spins \cite{Cucchietti05}. Equation  (\ref{gauss}) should be viewed as a second-quantized version of the same result. 

While the Gaussian decay of Eq. (\ref{gauss}) is irreversible, it is clear that due to the finite dimensionality of the system, coherence will revive. From Eq. (\ref{gauss_sum}), it is clear that $R_{12}(t+2\pi/U)=R_{12}(t)$ and that $R_{12}(t+\pi/U)=-R_{12}(t)$. Thus, relative-phase coherence with a $\pi$ shift should be restored after the revival time of $\tau_r=\pi/U$. Such collapse and revival was observed in interference experiments in optical lattices \cite{Greiner02}, where coherent states prepared in the Rabi regime, were allowed to evolve after the lattice height was raised, thereby afflicting a sudden transition to the Fock regime.

Summarizing this section, the dynamics of single particle coherence starting from the coherent state $|\pi/2,\pi\rangle$ in the absence of noise, depends on the interaction strength. In the Rabi regime phase is locked and the fringe-visibility does not decay but carries out small oscillations. For stronger interactions the fringe visibility is lost on a timescale which increases as the classical limit is approached by increasing $N$ while keeping $\kappa$ fixed, as $\log{N}$ in the Josephson regime and as $\sqrt{N}$ in the Fock regime.

\section{Decoherence - quantum Zeno suppression of phase diffusion}

Having established the different phase-diffusion regimes, we proceed to study the effect of noise on this process. We repeat the numerical calculations of the previous section, using the master equation (\ref{masterelement}) with non-vanishing values of the decoherence rates $\gamma_x$ and $\gamma_z$. Single-particle coherence is evaluated here using the von Neumann entropy of the reduced SPDM,
\begin{equation}
\label{entropy}
S\equiv -Tr(R\ln{R})=-\frac{1}{2}\ln{\left[\frac{(1+|\mathbf{s}|)^{(1+|\mathbf{s}|)}\cdot (1-|\mathbf{s}|)^{(1-|\mathbf{s}|)}}{4}\right]}~.
\end{equation} 

\begin{figure}
\centering
\includegraphics[width=0.5\textwidth] {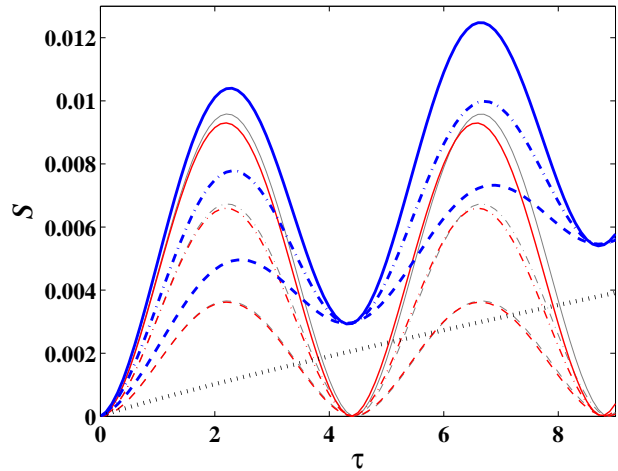}
\caption{(color online) Single-particle von Neumann entropy as a function of the rescaled time, for $\kappa=0.5$ (weak interactions). Initial conditions as in Fig.~\ref{qzefig2}. Solid, dash-dotted, and dashed lines correspond to numerically exact calculations  with $N = 100, 150, 300$ particles respectively. Non-Hamiltonian time evolution with $\gamma_z=10^{-4}, \gamma_x=0$ (bold blue lines) is compared with the Hamiltonian evolution with $\gamma_z=\gamma_x=0$ (normal red lines) and to mean-field decoherence with $\gamma_z=10^{-4}, \gamma_x=0$ (dotted black line). The corresponding BBR approximations for the Hamiltonian case, are shown in gray.} 
\label{qzefig5}
\end{figure}

\begin{figure}
\centering
\includegraphics[width=0.5\textwidth] {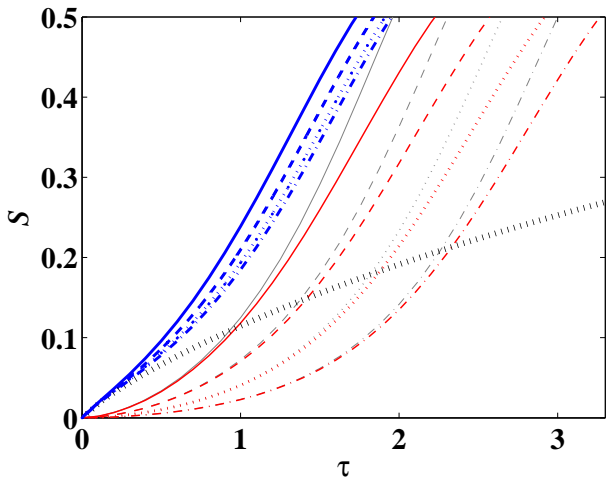}
\caption{(color online) Single particle von Neumann entropy as a function of rescaled time, for $\kappa=2$ and $\gamma_x=0$. Initial conditions as in Fig.~\ref{qzefig2}. Solid, dashed, dotted, and dash-dotted lines correspond to $N = 50, 100, 200, 400$ particles respectively. Non-Hamiltonian time evolution with $\gamma_z=0.1$ (bold blue lines) is compared to the Hamiltonian evolution with $\gamma_z=0$ (normal red lines) and to the mean-field decoherence with $\gamma_z=0.1$ (bold dotted black line). BBR approximate solutions for $\gamma_z=\gamma_x=0$, are shown in gray.} 
\label{qzefig6}
\end{figure}

The effect of local noise induced by collisions with thermal-cloud atoms, has been considered in previous work \cite{Vardi01,Anglin01}. From section III, it is evident that the pointer states for inelastic collisions within the BEC, are also the $\Lz$ eigenstates $|m\rangle$. Thus, since both entanglement and decoherence drive the system towards the same many-body states (see Fig.~\ref{qzefig1}(a)),  such decoherence {\it enhances} the deviation from classicality. This effect in the Rabi regime and in the Josephson regime, is demonstrated in Fig.~\ref{qzefig5} and Fig.~\ref{qzefig6}  respectively, where the dynamics of single-particle coherence in the presence of local noise $\gamma_z>0,\gamma_x=0$, is compared to its evolution when $\gamma_z=\gamma_x=0$ (the BBR approximate calculations of the decoherence-free, Hamiltonian evolution are shown in gray). 

\begin{figure}
\centering
\includegraphics[width=0.5\textwidth] {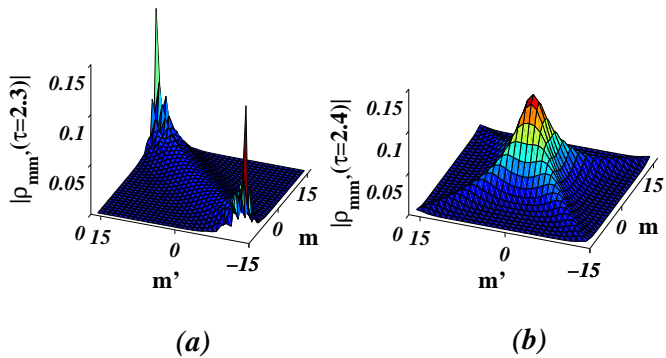}
\caption[Distribution of density matrices for dynamics with quantum noise]{Absolute values of the $N$-particle density matrix entries in relative Fock basis, for exact evolution with $N=30$ and $\kappa=2$, with noise. (a) The dephased cat state, created by local noise at $\tau=2.3$, for $\gamma_z=0.05, \gamma_x=0$, (b) The cat state is not formed and the one-body coherence is preserved by non-local noise at $\tau=2.4$, for $\gamma_z=0, \gamma_x=1$.} 
\label{qzefig7}
\end{figure} 

\begin{figure}
\centering
\includegraphics[width=0.5\textwidth] {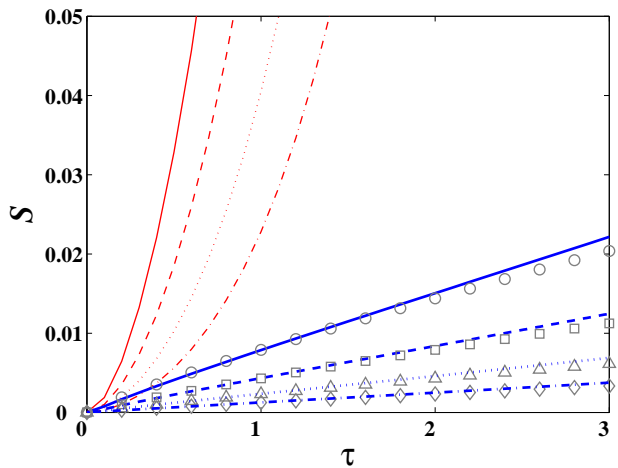}
\caption{(color online) Single particle von Neumann entropy as a function of rescaled time, for $\kappa=2$ and $\gamma_z=0$. Initial conditions as in Fig.~\ref{qzefig2}. Solid, dashed, dotted, and dash-dotted lines correspond to $N = 50, 100, 200, 400$ particles respectively. Non-Hamiltonian time evolution with $\gamma_x=10$ (bold blue lines) is compared to the Hamiltonian evolution with $\gamma_x=0$ (normal red lines). Anticipated QZE behavior is depicted by gray circles, squares, triangles, and diamonds. QZE is more pronounced for higher particle number.} 
\label{qzefig11}
\end{figure}

While  single-particle coherence is not conserved even in the classical limit because $\gamma_z\ne 0$, the deviation from mean-field theory in the Rabi regime increases due to decoherence (e.g. the minima of the single-particle entropy oscillations in Fig.~\ref{qzefig5}  lie above the mean-field entropy). Hence, phase-diffusion and local noise bootstrap to give faster dephasing. This trend is even more significant in the Josephson regime (Fig.~\ref{qzefig6}). The calculations in Fig.~\ref{qzefig6} differ from Ref. \cite{Vardi01,Anglin01}, only in the choice of the initial state, which was taken here to be $|\pi/2,\pi\rangle$ as compared to  $|0,0\rangle$ in \cite{Vardi01,Anglin01}, but since starting from the non-stationary $|0,0\rangle$ state drives the system towards $|\pi/2,\pi\rangle$, where most dephasing takes place, the results are qualitatively similar. The main conclusion remains that for strong interactions (see also bold blue lines in Fig.~\ref{qzefig11}) the quantum breaktime with $\gamma_z>0$ saturates to a finite value as compared to its logarithmic growth in agreement with Eq. (\ref{Runstable}), when  $\gamma_z=0$. 

The saturation of the dephasing time in the Josephson regime results from the known sensitivity of the dynamically-produced macroscopic superposition state to local noise \cite{Louis01,Micheli03,Huang06}.   The energy spacing between even- and odd cat states is exponentially small in $\kappa$, so that any small width incurred to these levels due to the local noise, may couple them and destroy the macroscopic coherence between the sites. The decohered state, shown in Fig.~\ref{qzefig7}(a), is close to the mixture$(|m{=}l\rangle\langle m{=}l| + |m{=}-l\rangle\langle m{=}-l|)/2$, corresponding to a 50\% probability of finding the particles in either site \cite{Louis01}, and to the origin ${\bf s}=(0,0,0)$ of the Bloch sphere. 

While local noise enhances the dephasing between the two modes, a different behavior may be anticipated if we consider other forms of decoherence. The multi-shot fringe visibility is proportional to the projection of the Bloch vector onto the $s_x s_y$ plane. Since ${\bf s}$ lies parallel to the $s_x$ axis, we have $R_{12}=s_x/2$. From Eq. (\ref{Rstable}) and Eq. (\ref{Runstable}) we see that for all interaction regimes this projection at $\tau< 1/\lambda$ scales {\it quadratically } rather than linearly in time,
\begin{equation}
|{\bf s}|=2|R_{12}|=\frac{2\left\langle\Lx\right\rangle}{N}=1-(\omega\tau)^2,
\end{equation}
where $\omega=\sqrt{2/N}|\coth(2i\Theta)|\lambda$. Thus, like the decay rate of any quantum-mechanical observable, the phase-diffusion rate vanishes at short times. Therefore, the rapid phase-diffusion of the initial unstable state in the Josephson and Fock regimes, may be {\it inhibited} by a quantum Zeno effect \cite{Khalfin68,Misra77,Itano90,Kofman00,Kofman01,Streed06,Haroche06}. In its discrete form, frequent projective measurements of the relative-number between the odd- and even quasi-momentum modes, may be carried out by addressing the overlap region of the two mode wavefunctions, separating the vanishing odd superposition from the non-vanishing even combination (see Fig.~\ref{qzefig1}). If these projections are taken at sufficiently short intervals with respect to the $\tau_c=1/\lambda$ correlation time $\delta{\tau} \ll \tau_c< 1/\omega$, they lead to exponential decay of the form,
\begin{equation}
|{\bf s}(\tau)|=\left[1-(\omega \delta\tau)^2\right]^j
\approx e^{\left[-j(\omega\delta\tau)^2\right]}
=e^{\left[-(\omega^2\delta\tau) \tau\right]}~.
\end{equation}
with $j$ being the number of successive measurements and $j\delta\tau=\tau$. The effective characteristic break time in the presence of decoherence
\begin{equation}
{\bar \tau}_{br}=\frac{1}{\omega^2\delta\tau}=\frac{N\tanh^2(2i\Theta)}
{2\lambda^2\delta\tau}\gg\frac{1}{\omega}~,
\label{breakzeno}
\end{equation}
should be compared in the Josephson regime with the decoherence-free breaktime,
\begin{equation}
\tau_{br}=\frac{\log\left[N\tanh^2(2i\Theta)/4\right]}{2\lambda}~,
\label{breaklog}
\end{equation}
or in the Fock regime with the $\tau_{br}=\sqrt{N}/\kappa$ characteristic time. Thus, on top of the standard extension of the decay time by frequent measurements, manifested by the comparison of the denominators in Eq. (\ref{breakzeno}) and Eq. (\ref{breaklog}) with $\lambda\delta\tau\ll 1$, there is a bosonic many-body factor of order $N/\log{N}\gg 1$ in the Josephson regime and of order $\sqrt{N}$ in the Fock regime. This factor results from the stabilization of the initial state which transforms the scaling of the breaktime with $N$ from logarithmic to linear. The many-body QZE will thus be significantly enhanced by bosonic stimulation.

The continuous version of the many-boson QZE amounts to the inclusion of the noise term proportional to $\gamma_x$. Provided that $\gamma_x\gg\lambda>\omega$, we can adiabatically eliminate $\Dyz$ in Eqs. (\ref{BBRlin}) and obtain,
\begin{equation}
|{\bf s}(\tau)|=\exp\left(-\frac{\omega^2}{2\gamma_x}\tau\right)~,
\label{scontzeno}
\end{equation}
so that the characteristic depletion time is 
\begin{equation}
\tilde{\tau}_{br}=\frac{2\gamma_x}{\omega^2}\gg \frac{1}{\omega}\gg\tau_{br}.
\end{equation} 

\begin{figure}
\centering
\includegraphics[width=0.5\textwidth] {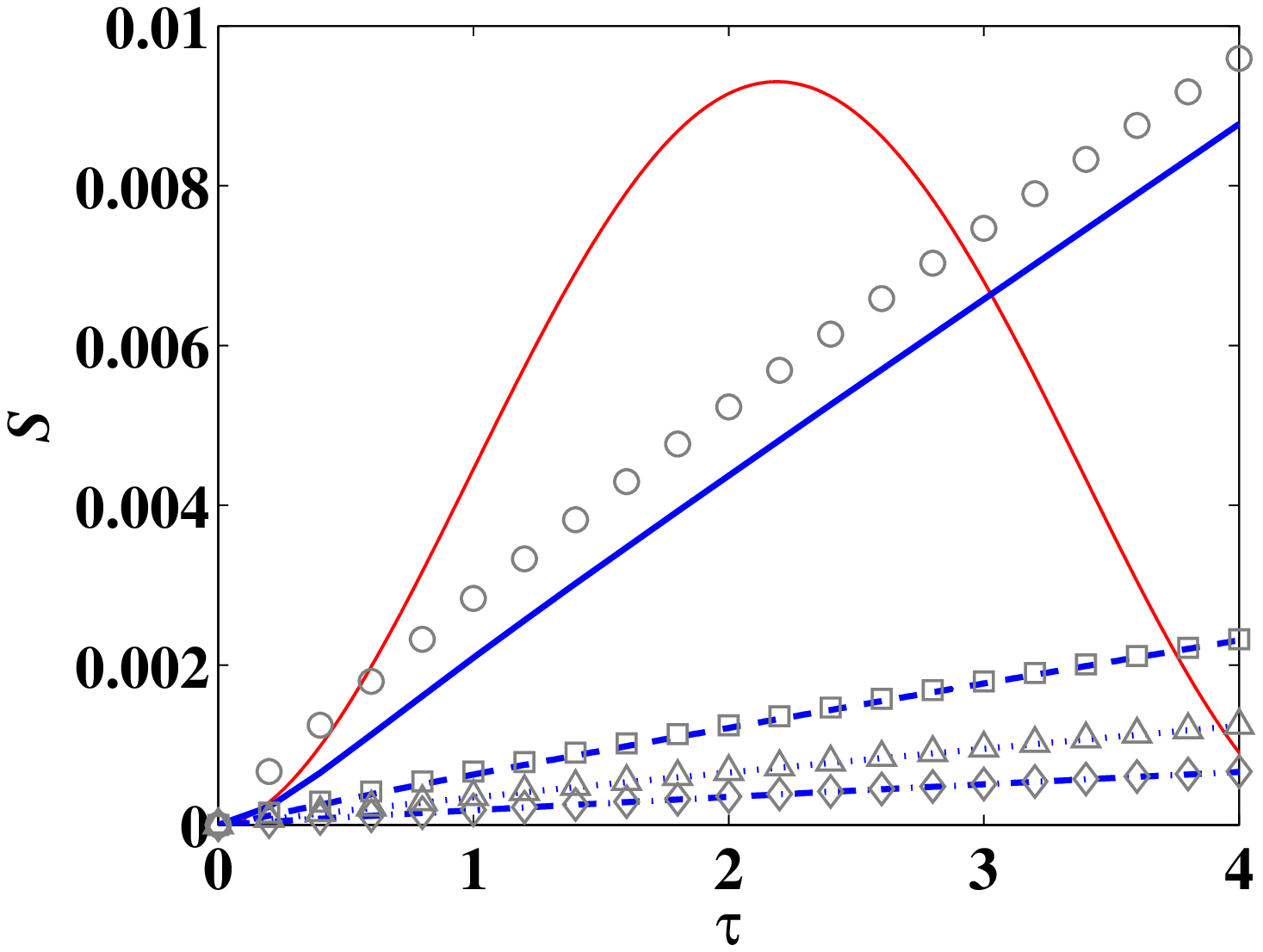}
\caption{(color online) Single particle von Neumann entropy as a function of rescaled time, for $\kappa=0.5$, $\gamma_z=0$, and $N=100$. Initial conditions as in Fig.~\ref{qzefig2}. The Hamiltonian dynamics with $\gamma_x=0$ (solid red) is compared to the evolution with non-vanishing $\gamma_x$ (bold blue lines).  Solid, dashed, dotted, and dash-dotted bold blue lines correspond to $\gamma_x=1,5,10,20$. Gray circles, squares, triangles, and diamonds respectively, portray the quantum-Zeno prediction of Eq. (\ref{scontzeno}), substituted into Eq. (\ref{entropy}). Analytical values coincide with numerical results for $\gamma\gg\lambda$.} 
\label{qzefig8}
\end{figure}

\begin{figure}
\centering
\includegraphics[width=0.5\textwidth] {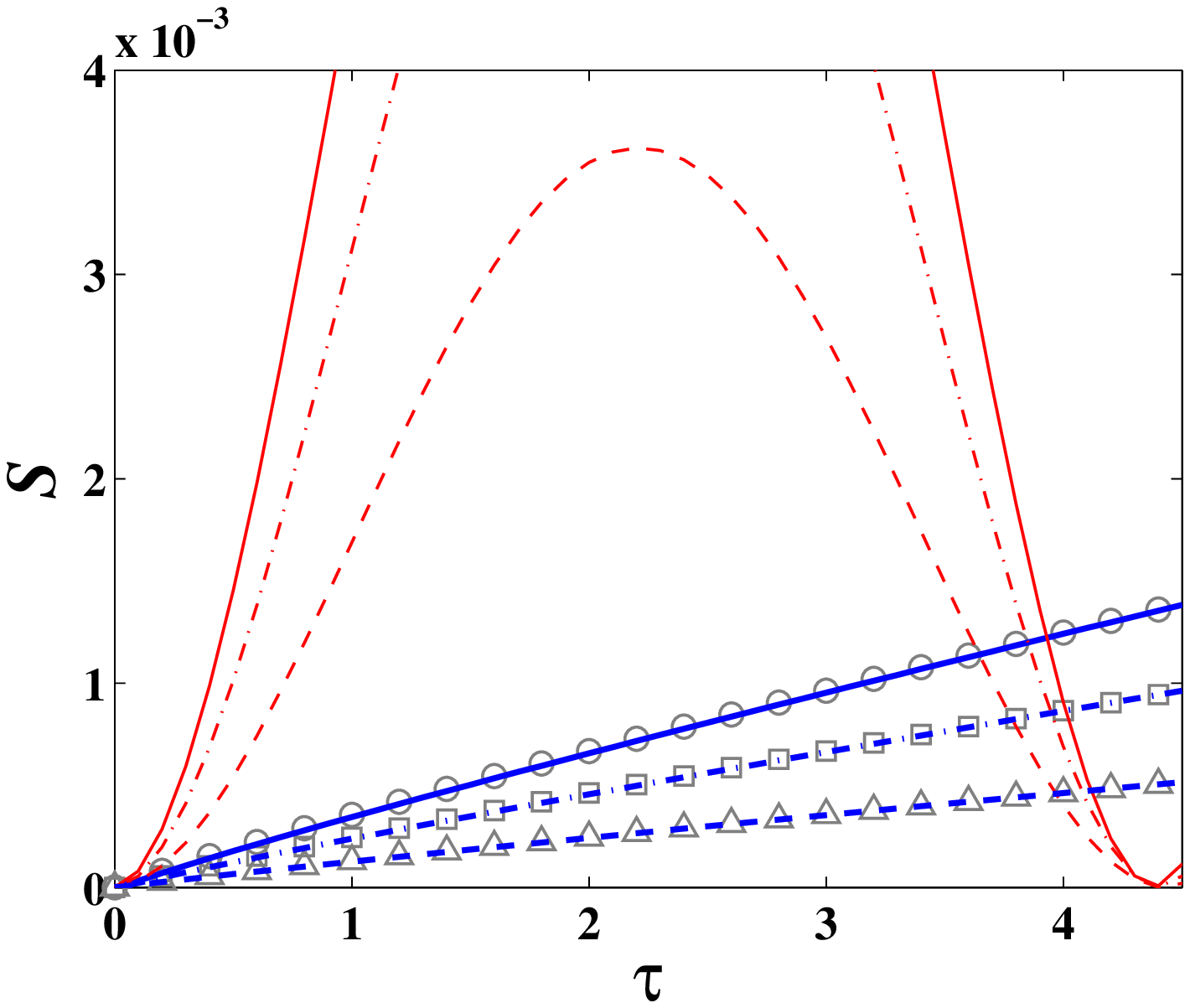}
\caption{(color online) Single particle von Neumann entropy as a function of rescaled time, for $\kappa=0.5$ and $\gamma_z=0$. Initial conditions as in Fig.~\ref{qzefig2}. Solid, dash-dotted, and dashed lines correspond to $N = 100, 150, 300$ particles respectively. Non-Hamiltonian time evolution with $\gamma_x=10$ (bold blue lines) is compared to the Hamiltonian evolution with $\gamma_x=0$ (normal red lines). Anticipated QZE behavior for the same particle numbers, is depicted by gray circles, squares, and triangles respectively.} 
\label{qzefig9}
\end{figure}

\begin{figure}
\centering
\includegraphics[width=0.5\textwidth] {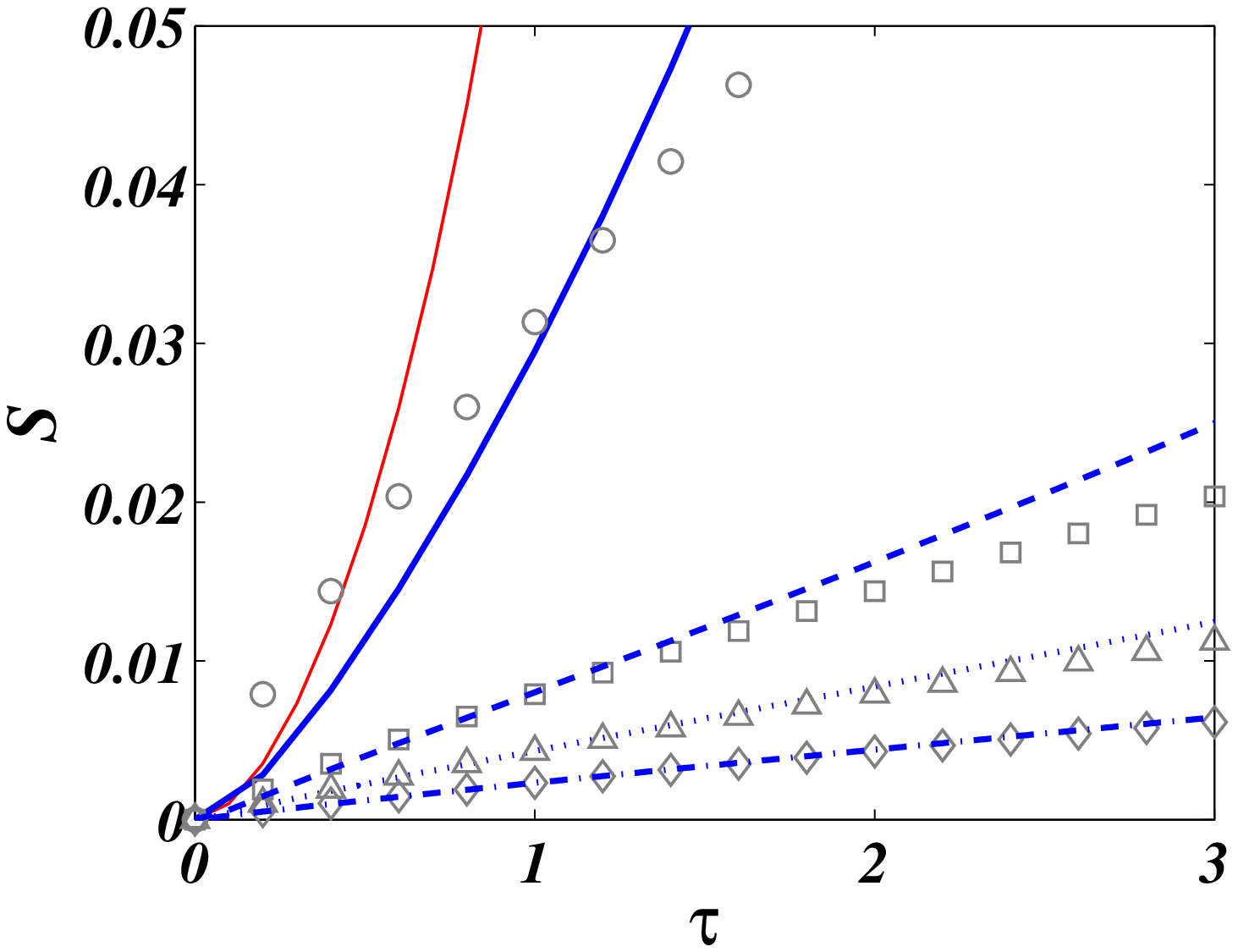}
\caption{(color online) Single particle von Neumann entropy as a function of rescaled time, for $\kappa=2$, $\gamma_z=0$, and $N=100$. Initial conditions as in Fig.~\ref{qzefig2}. The Hamiltonian dynamics with $\gamma_x=0$ (solid red) is compared to the evolution with non-vanishing $\gamma_x$ (bold blue lines).  Solid, dashed, dotted, and dash-dotted lines correspond to $\gamma_x=1,5,10,20$. Gray circles, squares, triangles, and diamonds respectively, portray the quantum-Zeno behavior of Eq. (\ref{scontzeno}) for the same values of $\gamma_x$, substituted into Eq. (\ref{entropy}). Similar to the weak-interactions case (Fig.~\ref{qzefig8}) good agreement with the QZE prediction is attained for $\gamma\gg\lambda$. A clear transition from initially quadratic to initially linear depletion is observed.} 
\label{qzefig10}
\end{figure}

The QZE suppressed phase-diffusion is illustrated in the numerical results of Fig.~\ref{qzefig8} and Fig.~\ref{qzefig9} for the Rabi regime and in Fig.~\ref{qzefig10} and Fig.~\ref{qzefig11} for the Josephson regime. In Fig.~\ref{qzefig8} and Fig.~\ref{qzefig10}, the initial evolution of the single particle entropy at large $\gamma_x$, is shown to approach the predicted value, obtained by substitution of Eq. (\ref{scontzeno}) into Eq. (\ref{entropy}). Both the weak-interaction oscillations and the strong-interaction phase-diffusion are replaced by an exponential (i.e. initially linear) decay, with a rate proportional to $1/(N\gamma_x)$, because they are piecewise constructed from similar short-time building blocks. The continuous projection onto the $\Lx$ axis, may thus either destabilize the stable excited state in the Rabi regime (Fig.~\ref{qzefig8}), or halt its unstable decay in the Josephson regime (Fig.~\ref{qzefig10}). From a quantum-to-classical transition perspective, when applied to the strongly-interacting system, this form of decoherence, which is tantamount to the continuous measurement of the relative-number in the rotated basis set of quasi-momentum states, will prolong the characteristic depletion time, thus restoring mean-field classicality. 

Bose enhancement of the QZE is illustrated in Fig.~\ref{qzefig9} and Fig.~\ref{qzefig10}, showing for weak- and strong coupling respectively, the QZE with varying particle numbers. Since the decay rates and the initial slopes are indeed dependent on $1/N$, it is evident that the transition from logarithmically growing breaktimes followed by an $N$-independent dephasing rate (normal red lines in Fig.~\ref{qzefig11}) to $1/N$ dependent decay rates (bold blue lines), implies the enhancement of the QZE by the expected factor, with stronger suppression of phase-diffusion for larger values of $N$. 

Looking at the full density matrix (Fig.~\ref{qzefig7}(b)), we see that the site-indiscriminate noise protects the initial coherent state, retaining a localized distribution somewhat smeared along the main and secondary diagonals. Thus, in contrast to the local noise which figuratively speaking 'kills the cat', the projection onto $L_x$ aborts the cat's very birth. Also by contrast, while even weak site-specific noise suffices to destroy the fragile coherence of the macroscopic superposition, the parity/momentum-specific noise needs to be sufficiently strong to induce the QZE effect.
 
\begin{figure}
\centering
\includegraphics[width=0.5\textwidth] {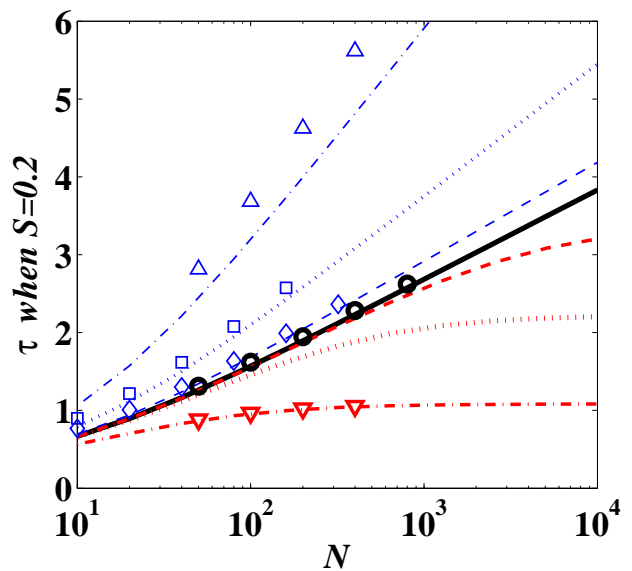}
\caption{(color online) Time at which $S$ reaches 0.2, starting from the coherent state $|\pi/2,\pi\rangle$, as a function of the particle number $N$, for $\kappa=2$. The bold solid black line, calculated from the BBR equations (\ref{BBR}), depicts the $\log N$ dependence for $\gamma_x=\gamma_z=0$, in good agreement with the exact numerical results (black circles). Bold red lines correspond to BBR calculations with $\gamma_x=0$ and $\gamma_z=5\times 10^{-4}$ (dashed), $5\times 10^{-3}$ (dotted), $5\times 10^{-2}$ (dash-dotted with inverted triangles denoting exact numerical results), all showing slower than logarithmic convergence with saturation to a finite breaktime at large $N$. The QZE transition to linear convergence is illustrated by the normal blue lines (BBR) and symbols (exact numerics), corresponding to $\gamma_z=0$ and $\gamma_x=0.1$ (dashed, diamonds), $0.4$ (dotted, squares), and $1$ (dash-dotted, triangles).} 
\label{qzefig12}
\end{figure}

To summarize our results, the quantum breaktime growth with increasing $N$ (keeping $\kappa=2$ fixed in the Josephson regime), is plotted in Fig.~\ref{qzefig12}. In the absence of decoherence, we obtain the expected $\log N$ convergence depicted by the bold black line (BBR) and circles (exact numerics, in good agreement with BBR). When local on-site noise ($\gamma_z>0$, bold red lines and symbols) is applied, this convergence is slowed down as shown earlier in Fig.~\ref{qzefig6} and the quantum breaktime saturates to a finite value which decreases logarithmically with increasing $\gamma_z$ \cite{Vardi01,Anglin01}. In comparison, when the noise is focused in the overlap region between the single-particle site-mode functions so as to measure the number-difference between their nonlocal odd- and even superposition ($\gamma_x>0$, normal blue lines and symbols), we observe faster than logarithmic convergence, approaching the QZE linear dependence on $N$ for large $\gamma_x$.

\section{Conclusions}

Using the two-site Bose-Hubbard model, we have studied the collisional phase-diffusion of the excited quasi-momentum number state with all particles in the odd superposition of the two modes. The loss of single-particle coherence is associated with classical stability, with bound oscillations in the stable Rabi regime and rapid depletion for strong interactions, in agreement with \cite{Vardi01,Anglin01}. The strong-coupling instability is presumably associated with the rapid heating observed in the merging of two condensates with a $\pi$ relative-phase, on an atom chip \cite{JoChoi07}. Linearizing the BBR equations, we obtained exact forms which describe the initial BEC depletion with good accuracy. The loss of coherence, attributed to the bose-amplification of initial spontaneous emission noise from the excited odd-superposition state to the even ground state, was shown to be quadratic (i.e. non-exponential) in $\tau$ within the $\tau_c=1/\lambda$ correlation time.

In order to study the effect of decoherence on phase-diffusion, we have considered two types of noise, corresponding to the measurement of the relative-particle number in two different basis sets. In the double-well realization, relative population imbalance between the site-modes can be measured by local noise on either site, whereas the relative number difference between the quasi-momentum modes (i.e. the even- and odd superpositions of the site-modes) can be determined by probing the overlap region in between the sites. The first type of decoherence was shown to accelerate depletion, in accordance with previous results \cite{Vardi01,Anglin01}. By contrast, odd-even relative number measurement, results in the suppression of collisional dephasing. This somewhat counterintuitive behavior, where one form of dephasing effectively suppresses another, is essentially a quantum Zeno effect, wherein the decay of an unstable state is suppressed by its continuous observation. However, unlike the classic QZE suppression of spontaneous decay \cite{Misra77}, the bosonic enhancement of the initial quantum noise implies that in subjecting the system to decoherence, a transition is made between $N$-independent decay rates, to rates that scale linearly with $N$. Consequently, the deceleration of phase-diffusion becomes more effective as the number of particles is increased by a $N/\log{N}$ factor (Josephson regime) or by a $\sqrt{N}$ factor (Fock regime), which may become very significant for realistic particle numbers in dilute gas BEC experiments.

In addition to the realization of the two-mode dynamics in the splitting and merging of atomic BECs \cite{Schumm05,Hofferberth07,Shin05,JoChoi07,Albiez05,ShinJo05,Gati06,Gati07,Levy07,Anderlini06,Sebby-Strabley07,Folling07}, similar behavior should be obtained in a wide range of many-boson quantum systems, which exhibit classical dynamical instabilities. One such example is the stimulated dissociation of a molecular BEC into bosonic fragments \cite{VardiYurovsky01,Vardi02,Moore02,TikhonenkovVardi07,Kheruntsyan02,Kheruntsyan05,Savage07}, where the short-time dynamics of the atom-molecule number difference depict a similar amplification of spontaneously-emitted atom-pairs, and therefore adhere to the same non-exponential behavior \cite{VardiYurovsky01,Kheruntsyan05}. Frequent or continuous observation of the relative atom-molecule number, will hence suppress this stimulated process, with similar bosonic enhancement factors.

\begin{acknowledgments}
This work was supported by the Israel Science Foundation (Grant number 582/07).
\end{acknowledgments}

\end{document}